\documentstyle[12pt,epsf]{article}

\newcommand{\ct}{\cite}
\newcommand{\lb}{\label}

\newcommand{\bc}{\begin{center}}
\newcommand{\ec}{\end{center}}
\newcommand{\bd}{\begin{displaymath}}
\newcommand{\ed}{\end{displaymath}}
\newcommand{\be}{\begin{equation}}
\newcommand{\ee}{\end{equation}}
\newcommand{\ba}{\begin{array}}
\newcommand{\ea}{\end{array}}
\newcommand{\bt}{\begin{tabular}}
\newcommand{\et}{\end{tabular}}
\newcommand{\un}{\underline}
\newcommand{\ov}{\overline}
\newcommand{\bp}{\begin{picture}}
\newcommand{\ep}{\end{picture}}
\newcommand{\bfi}{\begin{figure}}
\newcommand{\efi}{\end{figure}}

\def\fun#1#2{\lower3.6pt\vbox{\baselineskip0pt\lineskip.9pt
\ialign{$\mathsurround=0pt#1\hfil##\hfil$\crcr#2\crcr\sim\crcr}}}

\begin{document}

\vspace{2.5cm}

\title{\Large\bf {Anti-Grand Unification and the Phase Transitions
at the Planck Scale in Gauge Theories}}
\author{{\bf L.V.Laperashvili}
\footnote{{\bf E-mail}:laper@heron.itep.ru, larisa@vxitep.itep.ru}\\
\it Institute of Theoretical and Experimental Physics,\\
\it B.Cheremushkinskaya 25, 117218 Moscow, Russia}

\maketitle

\bc
{\bf Invited talk on the International Symposium\\
 "Frontiers of Fundamental Physics",\\
December 09-13, 2000, Hyderabad, India}
\ec

\pagenumbering{arabic}

\vspace{0.5cm}
The way of the Grand unification of all interactions and the
role of supersymmetry in GUTs are the problems of paramount
importance in the contemporary elementary particle physics. However,
at present time experiment doesn't indicate any manifestation of
the Supersymmetry (see reviews [1-3]).

In this connection, the Anti--Grand Unification Theory (AGUT) was
developed as a realistic alternative to SUSY GUTs by H.B.Nielsen (Niels
Bohr Insitute, Denmark) and his collaborators: D.L.Bennett,
C.D.Froggatt, L.V.Laperashvili, I.Picek and Y.Takanishi \ct{4}-\ct{17}.
According to the AGUT, the supersymmetry doesn't come into existence up
to the Planck energy scale:
\be
	  \mu_{Pl} = 1.2\cdot 10^{19}\;{\mbox GeV}.            \lb{1}
\ee
The SM is based on the group:
\be
   SMG = SU(3)_c \otimes SU(2)_L \otimes U(1)_Y.             \lb{2}
\ee
The AGUT suggests that at the scale $\mu_G\stackrel{<}{\sim} \mu_{Pl}$
there exists the more fundamental group G containing $N_{gen}$
copies of the Standard Model Group SMG:
\be
   G = SMG_1\otimes SMG_2\otimes ...\otimes SMG_{N_{gen}}
     \equiv {(SMG)}^{N_{gen}}
                                            \lb{3}
\ee
where the integer $N_{gen}$ designates the number of quark and lepton
generations. If $N_{gen}=3$, then the fundamental gauge group G is:
\be
    G = {(SMG)}^3 = SMG_1 \otimes SMG_2 \otimes SMG_3,       \lb{4}
\ee
or the generalized ones:
\be
                 G_f = {(SMG)}^3\otimes U(1)_f,                 \lb{5}
\ee
which is suggested by the fitting of fermion masses existing in the SM
(see Refs. \ct{9}-\ct{12}) and
\be
      G_{ext} = {(SMG\otimes U(1)_{B-L})}^3,             \lb{6}
\ee
which takes into account the see--saw mechanism with the right--handed
neutrino \ct{13},\ct{14}, also gives the reasonable fitting of
the SM fermion masses and describes all neutrino experiments known today.

The group $G_f$ contains the following gauge fields:
$3\times 8 = 24$ gluons, $3\times 3 = 9$ W-bosons and $3\times 1 + 1 = 4$
abelian gauge bosons.

At first sight, this ${(SMG)}^3\times U(1)_f$ group with its 37 generators
seems to be just one among many possible SM gauge group extensions.
However, it is not such an arbitrary choice. There are at least
reasonable requirements (postulates) on the gauge group G (or $G_f$,
or $G_{ext}$) which have uniquely to specify this group.  It should obey the
following postulates (the first two are also valid for SU(5) GUT):

\vspace{0.1cm}

1. G or $G_f$ should only contain transformation transforming the
known 45 Weyl fermions ( = 3 generations of 15 Weyl particles each)
-- counted as left handed, say -- into each other unitarily,
so that G (or $G_f$) must be a subgroup of U(45): $G\subseteq U(45)$.

\vspace{0.1cm}

2. No anomalies, neither gauge nor mixed. AGUT assumes that only
straightforward anomaly cancellation takes place and forbids the
Green-Schwarz type anomaly cancellation \ct{18}.

\vspace{0.1cm}

3. AGUT should NOT UNIFY the irreducible representations under the SM
gauge group, called here SMG (see Eq.(\ref{2})).

\vspace{0.1cm}

4. G is the maximal group satisfying the above-mentioned postulates.

\vspace{0.1cm}

There are five Higgs fields named $\phi_{WS}$, S, W, T, $\xi$
in the extended $G_f$--AGUT by Froggatt and Nielsen \ct{11},\ct{12}.
These fields break the AGUT to the SM what means that their vacuum expectation
values (VEV) are active. The field $\phi_{WS}$ corresponds
to the Weinberg---Salam theory, $<S>=1$, so that we have only
three free parameters --- three VEVs $\;<W>, <T>$ and $<\xi>$ to fit
the experiment in the framework of this model.
The authors of Refs.\ct{11},\ct{12} used them with
aim to find the best fit to conventional experimental data
for all fermion masses and mixing angles in the SM (see Table I).
The result presented by Table I is encouraging. The fit is given by the
${\chi}^2$ function (called here ${\tilde \chi}^2$). The lowest value
of ${\tilde \chi}^2 (\approx 1.87)$ gives the following VEVs:
\be
<S>=1;\quad\quad<W>=0.179;\quad\quad <T>=0.071;\quad\quad <\xi >=0.099.
                                                       \lb{7}
\ee
The extended Anti--GUT theory by Nielsen and Takanishi \ct{13},\ct{14},
which is described by the group of symmetry $G_{ext}$ (see Eq.(6)),
was suggested with aim to explain the neutrino oscillations.
Introducing the right--handed neutrino in the model, the authors replaced the
assumption 1 and considered U(48) group instead of U(45), so that
$G_{ext}$ is a subgroup of U(48): $G_{ext}\subseteq U(48)$. This group
ends up having 7 Higgs fields falling into 4 classes according to the order
of magnitude of the expectation values:

\vspace{0.1cm}

1) The smallest VEV Higgs field plays role of the SM Weinberg--Salam
Higgs field $\phi_{WS}$ having the weak scale value $<\phi_{WS}>=
246~GeV/{\sqrt 2}$.

\vspace{0.1cm}

2)The next smallest VEV Higgs field breaks all families $U(1)_{(B-L)}~$ group,
which is broken at the see--saw scale. This VEV is $<\phi_{(B-L)}>
\sim 10^{12}$ GeV. Such a field is absent in the "old" extended AGUT.

\vspace{0.1cm}

3) The next 4 Higgs fields are W, T, $\xi$ and $\chi$, which have VEVs
of the order of a factor 10 to 50 under the Planck unit. That means
that if intermediate propagators have scales given by the Planck scale,
as it is assumed in the AGUT in general, then they will give rise to
suppression factors of the order 1/10 each time they are needed to cause
a transition. The field $\chi$ is absent in the "old" $G_f$--AGUT.
It was introduced in Refs.\ct{13},\ct{14} for the purpose of the study
neutrinos.

\vspace{0.1cm}

4) The last one, with VEV of the same order as the Planck scale,
is the Higgs field S. It had VEV $<S>=1$ in the "old" extended
AGUT by Froggatt and Nielsen (with $G_f$ group of symmetry), but this VEV
is not equal to unity in the "new" extended AGUT. Therefore there is
a possibility to observe phenomenological consequences of the field S
in the Nielsen--Takanishi model.

Typical fit to the masses and mixing angles for the SM leptons and
quarks in the framework of the $G_{ext}$--AGUT is given by Table II.
The lowest value of ${\tilde \chi}^2$ is $\approx 1.46$.
In contrast to the "old" extended AGUT, the new results are more
encouraging.

The AGUT approach is used in conjuction with the Multiple Point
Principle (MPP) proposed by D.L.Bennett and H.B.Nielsen \ct{6}.
According to this principle Nature seeks a special point --- the Multiple
Critical Point (MCP) --- which is a point on the phase diagram of the
fundamental regulirized gauge theory G (or $G_f$, or $G_{ext}$), where
the vacua of all fields existing in Nature are degenerate having the same
vacuum energy density. This is the Multiple Point Principle.
Such a phase diagram has axes given by all coupling constants
considered in theory. Then all (or just many) numbers of phases
meet at the MCP.

In the AGUT at some point $\mu_G$ the group G (or $G_f$, or $G_{ext}$)
undergoes spontaneous breakdown to the diagonal subgroup:
\be
     G \longrightarrow G_{diag.subgr.} = \{g,g,g|| g\in SMG\},
                                                            \lb{8}
\ee
which is identified with the usual (lowenergy) group SMG.

Multiple Point Model assumes the existence of MCP at the Planck scale,
insofar as gravity may be "critical" at the Planck scale.

The idea of MPP has its origin from the lattice investigations of
gauge theories. In particular, Monte Carlo simulations on lattice
of U(1)-, SU(2)- and SU(3)- gauge theories indicate the existence
of a triple (critical) point, which is a boundary point of three
first order phase transitions.
Using the Monte Carlo results on lattice, it is possible to make
theoretical calculations of the critical coupling constants and
obtain slightly more accurate predictions of the AGUT for the SM fine
structure constants.

In the SM the usual definition of coupling constants is used:
\be
   \alpha_1 = \frac 53 \frac{\alpha}{{\cos}^2 \Theta_{\ov{MS}}},\quad\quad
   \alpha_2 = \frac{\alpha}{{\sin}^2 \Theta_{\ov{MS}}},\quad\quad
   \alpha_3 \equiv \alpha_s =\frac{g^2_s}{4\pi },
                                                            \lb{9}
\ee
where $\alpha$ and $\alpha_s$ are the electromagnetic and strong fine
structure constants, respectively. All values
are defined in the Modified minimal substraction
scheme ($\ov{MS}$). Using experimentally given parameters and the
renormalization group equations (RGE), it is possible to extrapolate the
experimental values of three inverse running constants $\alpha_i^{-1}$
(here $i=1,2,3$ corresponds to U(1), SU(2) and SU(3) groups)
from Electroweak scale to the
Planck scale. The precision of the LEP data allows to make this
extrapolation with small errors. Assuming that the RGEs
are contingent not encountering new particles up to
$\mu \stackrel{<}{\sim}\mu_{Pl}$ and doing the extrapolation with one
Higgs doublet under the assumption of a "desert", the
following result for the inverses $\alpha^{-1}_{Y,2,3}$ ($\alpha_Y\equiv
\frac 35 \alpha_1$) was obtained in Ref.\ct{6}:
\be
  {\alpha}^{-1}_Y(\mu_{Pl})\approx 55.5;\quad\quad
  {\alpha}^{-1}_2(\mu_{Pl})\approx 49.5; \quad\quad
  {\alpha}^{-1}_3(\mu_{Pl})\approx 54.
                                           \lb{10}
\ee
The extrapolation of $\alpha_i^{-1}(\mu )$ up to the point $\mu=\mu_G$
is shown in Fig.1. The AGUT predicts their values at the scale
$\mu_G\sim 10^{18}$ GeV (which is very close to $\mu_{MCP}=\mu_{Pl}$)
in terms of the critical couplings $\alpha_{i,crit}$ taken from the
lattice gauge theory \ct{19}-\ct{23}:
\be
     \alpha_i(\mu_{Pl}) = \frac {\alpha_{i,crit}}{N_{gen}} =
          \frac{\alpha_{i,crit}}{3}
					\lb{11}
\ee
for i=2,3 and
\be
       \alpha_1(\mu_{Pl}) = \frac {\alpha_{1,crit}}{{\frac 12}N_{gen}
             (N_{gen} + 1)} = \frac{\alpha_{1,crit}}{6}
							  \lb{12}
\ee
for U(1).

According to the AGUT, at the Planck scale the running constants
$\alpha_1$ (or $\alpha_Y\equiv\frac{3}{5}\alpha_1$), $\alpha_2$ and
$\alpha_3$, as chosen by Nature, are just the ones corresponding to the MCP.

There exists a simple explanation of the relations (\ref{11}) and (\ref{12}).
As it was mentioned above, the group G breaks down
at $\mu=\mu_G$. It should be said that at
the very high energies $\mu \ge \mu_G \stackrel{<}{\sim}\mu_{Pl}$ (see Fig.1)
each generation has its own gluons, own W's etc. The breaking makes
only linear combination of a certain color combination of gluons which
exists below $\mu=\mu_G$ and down to the low energies.
We can say that the phenomenological gluon is a linear
combination (with amplitude $1/\sqrt 3$ for $N_{gen}=3$) for each of the
AGUT--gluons of the same color combination. This means that coupling constant
for the phenomenological gluon has a strength that is $\sqrt 3$ times smaller,
if as we effectively assume that three AGUT SU(3) couplings are equal
to each other.
Then we have the following formula connecting the fine structure constants
of G--theory (e.g. AGUT) and low energy surviving diagonal subgroup
$G_{\mbox{diag.subg.}}\subseteq {(SMG)}^3$ given by Eq.(\ref{8}):
\be
\alpha_{\mbox{diag},i}^{-1} = \alpha_{\mbox{1st gen.},i}^ {-1} +
\alpha_{\mbox{2nd gen.},i}^{-1} + \alpha_{\mbox{3rd gen.},i}^{-1}.
                                            \lb{13}
\ee
Here i = U(1), SU(2), SU(3), and i=3 means that we talk about the gluon
couplings.
For non--Abelian theories we immediately obtain Eq.(\ref{11}) from
Eq.(\ref{13}) at the critical point (MCP).

In contrast to non-Abelian theories, in which the gauge invariance
forbids the mixed (in generations) terms in the Lagrangian of
G--theory, the U(1)--sector of the AGUT contains such mixed
terms what explains the difference between the expressions
(\ref{11}) and (\ref{12}).

Using Monte Carlo results on lattice the AGUT predicts \ct{6} :
\be
\alpha_Y^{-1}(\mu_{MCP}) = 55 \pm 6,\quad
\alpha_2^{-1}(\mu_{MCP}) = 49.5 \pm 3, \quad
\alpha_3^{-1}(\mu_{MCP}) = 57 \pm 3,
                                        \lb{14}
\ee
in correspondence with the result (10).

According to Eq.(\ref{12}), the first values of Eqs.(\ref{10}) and (\ref{13})
gives the following estimation for the U(1) fine structure constant at
$\mu = \mu_{MCP}$:
\be
          \alpha_{crit}^{-1}\sim 9.                  \lb{15}
\ee
The Monte Carlo simulations of the lattice U(1) gauge theory gives
\ct{20}-\ct{22} :
\be
      \alpha_{crit}\approx 0.20\pm 0.015,\quad {\mbox {or}}\quad
      \alpha_{crit}^{-1}\approx 5.                  \lb{16}
\ee
However, it is possible to show (see \ct{6}) that quantum
fluctuations encrease the value of $\alpha_{crit}^{-1}~$ giving
the value (\ref{15}).

The following hypothesis was stated in Refs.\ct{16},\ct{17}: it is
possible that the existence of monopoles at superhigh energies
--- at a short distance from the Planck energy scale (1) --- plays
an essential role in the phase transitions at the Planck scale when
all fields existing in the SM turn into the new phase, say,
(super)string phase.
In the previous works \ct{6},\ct{7},\ct{15} the investigation
of the phase transition phenomena and, in particular, the
calculation of the U(1) critical coupling constant were connected with
the existence
of artifact monopoles in the lattice gauge theory and also in the Wilson
loop action model, which we proposed in Ref.\ct{15}.
Now, instead of using the lattice or Wilson loop cut-off, we are going to
introduce physically existing monopoles into the theory as fundamental fields.

Developing a version of the local field theory of the Higgs scalar
monopoles and electrically charged particles, we consider
an Abelian gauge theory in the Zwanziger formalism \ct{24}-\ct{28}
and look for a/or rather several phase transitions connected with
the monopoles forming a condensate in the vacuum.

The Zwanziger formalism \ct{24},\ct{25}
(see also \ct{26},\ct{27} and review \ct{28}) considers
two potentials $A_{\mu}(x)$ and $B_{\mu}(x)$ describing one physical
photon with two physical degrees of freedom.
Now and below we call this theory QEMD ("Quantum
ElectroMagnetoDynamics").

In QEMD the total field system of the gauge, electrically ($\Psi$)
and magnetically ($\Phi$) charged fields (with charges $e$ and $g$,
respectively) is described by the partition
function which has the following form in the Euclidean space:
\be
 Z = \int [DA][DB][D\Phi ][D\Phi^{+}][D\Psi ][D\Psi^{+}]e^{-S},
                                                   \lb{17}
\ee
where
\be
S = \int d^4x L(x) =
           S_{Zw}(A,B) + S_{gf} + S_{(matter)}.
                                          \lb{18}
\ee
The Zwanziger action $S_{Zw}(A,B)$ is given by:
$$
      S_{Zw}(A,B) = \int d^4x [\frac 12 {(n\cdot[\partial \wedge A])}^2 +
                  \frac 12 {(n\cdot[\partial \wedge B])}^2 +\\
$$
\be
     +\frac i2(n\cdot[\partial \wedge A])(n\cdot{[\partial \wedge B]}^*)
       - \frac i2(n\cdot[\partial \wedge B])(n\cdot{[\partial \wedge A]}^*)],
                                                 \lb{19}
\ee
where we have used the following designations:
$$
       {[A \wedge B]}_{\mu\nu} = A_{\mu}B_{\nu} - A_{\nu}B_{\mu},
\quad {(n\cdot[A \wedge B])}_{\mu} = n_{\nu}{(A \wedge B)}_{\nu\mu},\\
$$
\be
\quad {G}^*_{\mu\nu} = \frac 12\epsilon_{\mu\nu\lambda\rho}G_{\lambda\rho}.
                                   \lb{20}
\ee
In Eqs.(\ref{19}) and(\ref{20}) the unit vector $n_{\mu}$ represents
the fixed direction of the Dirac string in the 4--space.

The action $S_{(matter)} = \int d^4x L_{(matter)}(x)$
describes the electrically and magnetically charged matter
fields. $S_{gf}$ is the gauge--fixing action (see \ct{26}).

Let us consider now the Lagrangian $L_{(matter)}$ describing the Higgs
scalar fields $\Psi(x)$ and $\Phi(x)$ interacting with gauge fields
$A_{\mu}(x)$ and $B_{\mu}(x)$, respectively:
\be
   L_{(matter)}(x) = \frac 12 {|D_{\mu}\Psi|}^2 +
     \frac 12{|{\tilde D}_{\mu} \Phi |}^2 - U(\Psi, \Phi),
                                   \lb{21}
\ee
where
\be
       D_{\mu} = \partial_{\mu} - ieA_{\mu},\quad {\mbox{and}}\quad
       {\tilde D}_{\mu} = \partial_{\mu} - igB_{\mu}   \lb{23}
\ee
are covariant derivatives;
\be
U(\Psi, \Phi) = \frac 12 \mu_e^2{|\Psi|}^2 + \frac{\lambda_e}4{|\Psi|}^4
        + \frac 12 \mu_m^2{|\Phi|}^2 + \frac{\lambda_m}4{|\Phi|}^4
        + \lambda_1{|\Psi|}^2{|\Phi|}^2
                                           \lb{24}
\ee
is the Higgs potential for the electrically and magnetically charged
fields $\Psi$ and $\Phi$.
The complex scalar fields:
\be
\Psi = \psi + i\zeta \qquad \mbox{and}\qquad
\Phi = \phi + i\chi   \lb{25}
\ee
contain the Higgs $(\psi, \phi )$ and Goldstone $(\zeta, \chi )$ boson fields.

The Lorentz invariance is lost in the Zwanziger Lagrangian (\ref{19})
because of depending on a fixed vector $n_{\mu}$, but this invariance
regained for the quantized values of coupling constants $e$ and $g$
obeying the Dirac relation:
\be
                 e_ig_j = 2\pi n_{ij},~~~~~n_{ij} \in Z.
                                               \lb{26}
\ee
Considering the electric and magnetic fine structure constants:
{\large $\alpha =
\frac {\displaystyle{e^2}}{\displaystyle{4\pi}}$ and
$\tilde \alpha =
\frac {\displaystyle{g^2}}{\displaystyle{4\pi}}$}
we have the invariance of the QEMD under the interchange
         $ \alpha \leftrightarrow \tilde \alpha.$

For $n_{ij}=1$ from the Dirac relation (\ref{26}) we have:
\be
             \alpha \tilde \alpha = \frac 14.     \lb{27}
\ee
The effective potential in the Higgs model of electrodynamics for
a charged scalar field was calculated in the one-loop approximation for the
first time by the authors of Ref.\ct{29} (see also review \ct{30}).
Using this method we can construct the effective potential
(also in the one--loop approximation) for the theory described by the
partition function (\ref{17}) with the action $S$.

Let us consider now the shifts: $\Psi = \Psi_B + {\hat \Psi}(x), \quad
\Phi (x) = \Phi_B + {\hat \Phi}(x)$
with $\Psi_B$ and $\Phi_B$ as background fields and calculate the
following expression for the partition function in the one-loop
approximation:
$$
  Z = \int [DA][DB][D\hat \Phi][D{\hat \Phi}^{+}][D\hat \Psi]
      [D{\hat \Psi}^{+}]\times \\
$$
$$
   \exp\{ - S(A,B,\Phi_B,\Psi_B)
   - \int d^4x [\frac{\delta S(\Phi)}{\delta \Phi(x)}|_{\Phi=
   \Phi_B}{\hat \Phi}(x) +
   \frac{\delta S(\Psi)}{\delta \Psi(x)}|_{\Psi=\Psi_B}{\hat \Psi}(x)
                + h.c. ]\}\\
$$
\be
    =\exp\{ - F(\Psi_B, \Phi_B, e^2, g^2, \mu_e^2, \mu_m^2,
                                \lambda_e, \lambda_m)\}.
                                       \lb{29}
\ee

Using the representations (\ref{25}), we obtain the effective potential:
\be
  V_{eff} = F(\psi_B, \phi_B, e^2, g^2, \mu_e^2, \mu_m^2, \lambda_e,
                      \lambda_m)
                                  \lb{30}
\ee
given by the function $F$ of Eq.(\ref{29}) for the constant background
fields:$~\Psi_B = \psi_B = \mbox{const},\quad\Phi_B = \phi_B = \mbox{const}$.

The effective potential (\ref{30}) has several minima. Their position
depends on $e^2, g^2, \mu^2_{e,m}$ and $\lambda_{e,m}$.
If the first local minimum occurs at $\psi_B=0$ and
$\phi_B=0$, it corresponds to the the Coulomb-like phase in our description.

We are interested in the phase transition from the
Coulomb-like phase "$\psi_B = \phi_B = 0$" to the confinement phase
"$\psi_B = 0,\; \phi_B = \phi_0 \neq 0$". In this case the one--loop
effective potential for monopoles coincides with the expression
of the effective potential calculated by authors of Ref.\ct{29} for scalar
electrodynamics and extended to the massive theory in Ref.\ct{31}.

Assuming the existence of the first vacuum at $\phi_B=0$ and
using from now the designations:$\quad\mu = \mu_m,\quad\lambda = \lambda_m$,
we have the effective potential in the Higgs monopole model described
by the following expression equivalent to that considered in Ref.\ct{29}:
\be
V_{eff}(\phi^2)
= \frac{\mu^2_{run}}{2}\phi_B^2 + \frac{\lambda_{run}}{4}\phi_B^4
     + \frac{\mu^4}{64\pi^2}\log\frac{(\mu^2 + 3\lambda \phi_B^2)(\mu^2 +
       \lambda \phi_B^2)}{\mu^4},
                                            \lb{33}
\ee
where
\be
  \lambda_{run}(\phi_B^2)
   = \lambda + \frac{1}{16\pi^2} [ 3g^4\log \frac{\phi_B^2}{M^2}
   + 9{\lambda}^2\log\frac{\mu^2 + 3\lambda \phi_B^2}{M^2} +
     {\lambda}^2\log\frac{\mu^2 + \lambda\phi_B^2}{M^2}],
                                    \lb{34}
\ee
\be
   \mu^2_{run}(\phi_B^2)
   = \mu^2 + \frac{\lambda\mu^2}{16\pi^2}[ 3\log\frac{\mu^2 +
   3\lambda \phi_B^2}{M^2} + \log\frac{\mu^2 + \lambda\phi_B^2}{M^2}].
                                 \lb{35}
\ee
Here $M$ is the cut--off scale.

As it was shown in Ref.\ct{29}, the one--loop effective potential
(\ref{33}) can be improved by the consideration of the renormalization
group equation (RGE).
According to Refs.\ct{29}-\ct{31}, RGE for the
improved one--loop effective potential is given by the following
expression:
\be
 (M^2\frac{\partial}{\partial M^2} +
    \beta_{\lambda}\frac{\partial}{\partial \lambda} +
    \beta_g\frac{\partial}{\partial g} +
    \beta_{(\mu^2)}{\mu^2}\frac{\partial}{\partial \mu^2} -
    \gamma \phi^2 \frac{\partial}{\partial \phi^2}) V_{eff}(\phi^2) = 0,
                        \lb{36}
\ee
where the function $\gamma $ is the anomalous dimension:
     $ \gamma(\frac{\phi}M) = - \frac{\partial \phi}{\partial M}.$
The $\gamma$--expression for monopoles is given by Ref.\ct{29}
with replacement $e\rightarrow g$:
\be
          \gamma = - \frac{3g_{run}^2}{16\pi^2}.     \lb{42}
\ee

RGE (\ref{36}) leads to a new improved effective potential:
\be
   V_{eff}(\phi^2)
         = \frac 12 \mu^2_{run}(t) G^2(t)\phi^2 +
               \frac 14 \lambda_{run}(t) G^4(t) \phi^4,
                                           \lb{37}
\ee
where
\be
    G(t)\equiv \exp[ - \frac 12 \int_0^t dt' \gamma\biggl(g_{run}(t'),
         \lambda_{run}(t')\biggl)]
\quad {\mbox{with}} \quad
t = \log (\phi^2/{M^2}).            \lb{38}
\ee
Let us write now the one--loop potential (\ref{33}) as
\be
	 V_{eff} = V_0 + V_1,      \qquad
{\mbox{where}}\qquad
   V_0 = \frac{\mu^2}2 \phi^2 + \frac{\lambda}4 \phi^4, \lb{39}
\ee
$$
  V_1 = \frac{1}{64\pi^2}[ 3g^4 {\phi}^4\log\frac{\phi^2}{M^2}
+ {(\mu^2 + 3\lambda {\phi}^2)}^2\log\frac{\mu^2 + 3\lambda
\phi^2}{M^2}
$$
\be
   + {(\mu^2 +\lambda \phi^2)}^2\log\frac{\mu^2
+ \lambda \phi^2}{M^2} - 2\mu^4\log\frac{\mu^2}{M^2}].
                      \lb{40}
\ee
We can plug this $V_{eff}$ into RGE (\ref{36}) and obtain the
following RG--equations (see \ct{30}):
\be
\frac{d\lambda_{run}}{dt} =
 \frac 1{16\pi^2}( 3g^4_{run} +10 \lambda^2_{run} - 6\lambda_{run}g^2_{run}),
                                \lb{43}
\ee
\be
\frac{d\mu^2_{run}}{dt} = \frac{\mu^2_{run}}{16\pi^2}( 4\lambda_{run} -
			   3g^2_{run} ).
                                                 \lb{44}
\ee
The Dirac relation and the RGE
for the electric and magnetic fine structure constants $\alpha $
and $\tilde \alpha $ were investigated in detail in the
recent paper \ct{27}. The following result was obtained.

If we have the electrically and magnetically charged particles
existing simultaneously for $\tilde \mu > \tilde \mu_{(threshold)}$
and if in some region of $\tilde \mu$ their
$\beta$--functions are computable perturbatively as a power series in $e^2$
and $g^2$, then the Dirac relation is valid not only for the "bare"
elementary charges $e_0$ and $g_0$, but also for the
renormalized effective charges $e$ and $g$ (see \ct{32} and review \ct{28}),
and the following RGEs (obtained in Ref.\ct{27}) take place:
\be
  \frac{ d\log \alpha(p)}{dt} = -
  \frac{ d\log {\tilde \alpha}(p)}{dt} =
   \beta^{(e)}(\alpha ) - \beta^{(m)}(\tilde \alpha ).   \lb{45}
\ee
These RGEs are in accordance with the Dirac relation (\ref{27})
and the dual symmetry considered above.
By restricting ourselves to the two--loop approximation for $\beta$--
functions, we have the following equations (\ref{45}) for scalar particles:
\be
  \frac{ d\log \alpha(p)}{dt} = -
  \frac{ d\log {\tilde \alpha}(p)}{dt} =
   \frac{\alpha - \tilde \alpha }{12\pi}( 1 + 3\frac{\alpha
          + \tilde \alpha}{4\pi} + ....).
                                           \lb{46}
\ee
According to Eq.(\ref{46}), the two--loop contribution is
not more than $30\%$ if both $\alpha $ and
$\tilde {\alpha}$ obey the following requirement:
\be
  0.25 \stackrel{<}{\sim }{\large \alpha, \tilde{\alpha}}\stackrel{<}
                {\sim } 1.
                                        \lb{47}
\ee
The lattice simulations of compact QED give the behavior of
the effective fine structure constant $\alpha(\beta)$
($\beta=1/e_0^2$ , and $e_0$ is the bare electric charge)
in the vicinity of the phase transition point
(see Refs.\ct{20},\ct{22}).
The following critical values of the fine structure
constant $\alpha $ and $\tilde \alpha$ was obtained in Ref.\ct{22}:
\be
\alpha_{crit}^{lat}\approx{0.20 \pm 0.015}\quad\quad
{\tilde \alpha}_{crit}^{lat}\approx{1.25 \pm 0.10}\quad\quad
\mbox{at}\quad\quad \beta_{crit}\approx{1.011}.
                                             \lb{48}
\ee
These values almost coincide with the borders of the
requirement (\ref{47}) given by the perturbation theory for
$\beta$--functions.

In the one--loop approximation, we have:
\be
\frac{dg^2_{run}}{dt} = \frac{g^4_{run}}{48\pi^2} - \frac 1{12}.
                                             \lb{49}
\ee
Here the second term describes the influence of
the electrically charged fields on the behavior of the monopole charge.

Investigating the phase transition from
the Coulomb--like phase "$\psi_B = \phi_B = 0$" to the phase with
"$\psi_B = 0,\; \phi_B = \phi_0 \neq 0$", we see that the
effective potential (\ref{37}) has
the first and the second minima appearing at $\phi = 0$ and
$\phi = \phi_0$, respectively. They are shown in Fig.2 by the curve "1".
These minima of $V_{eff}(\phi^2)$ correspond to the different vacua
arising in the model.

The conditions for the existence of degenerate vacua are given by the
following requirements:
\be
          V_{eff}(0) = V_{eff}(\phi_0^2) = 0,     \lb{51}
\ee
\be
               V'_{eff}(\phi_0^2)
    \equiv \frac{\partial V_{eff}}{\partial \phi^2}|_{\phi=\phi_0} = 0,
                                        \lb{52}
\ee
\be
                V''_{eff}(\phi_0^2)
    \equiv \frac{\partial^2 V_{eff}}{\partial {(\phi^2)}^2}|_{\phi=\phi_0}
> 0.
                                            \lb{53}
\ee
From the first equation (\ref{51}) applied to Eq.(\ref{37}) we have:
\be
       \mu_{run}^2 = - \frac 12 \lambda_{run}(t_0){\phi_0^2}G^2(t_0),
\quad
{\mbox{where}}\quad t_0 = \log(\phi_0^2/M^2).
                                         \lb{54}
\ee
The joint solution of equations
$V_{eff}(\phi_0^2) = V'_{eff}(\phi_0^2) = 0 $ gives:
\be
     g^4_{crit} = - 2\lambda_{run}(\frac{8\pi^2}3 + \lambda_{run}).
                                            \lb{55}
\ee
The curve (\ref{55}) is represented on the phase diagram
$(\lambda_{run}; g^4_{run})$ of Fig.3 by the curve "1" which describes
a border between the "Coulomb--like" phase with $V_{eff} \ge 0$
and the confinement ones having $V_{eff}^{min} < 0$.

The next step is the calculation of the second derivative of the effective
potential.
Let us consider now the case when this second derivative
changes its sign giving a maximum of $V_{eff}$ instead of the minimum
at $\phi^2 = \phi_0^2$. Such a possibility is shown in Fig.2 by
the dashed curve "2".
Now two additional minima at $\phi^2 = \phi_1^2$ and $\phi^2 = \phi_2^2$
appear in our theory. They correspond to two different confinement phases
related with the confinement of the electrically  charged particles.
If these two minima are degenerate, then we have the following
requirements:
\be
       V_{eff}(\phi_1^2) = V_{eff}(\phi_2^2) < 0,\quad
        {V'}_{eff}(\phi_1^2) = {V'}_{eff}(\phi_2^2) = 0,
                                            \lb{57}
\ee
which describe the border between the confinement phases "conf.1" and "conf.2"
presented in Fig.3 by curve "3". This curve "3"
meets the curve "1" at the triple point A.
According to the illustration shown in Fig.3, it is obvious
that this triple point A is given by the following requirements:
\be
    V_{eff}(\phi_0^2) = V'_{eff}(\phi_0^2) = V''_{eff}(\phi_0^2) = 0.
                                       \lb{58}
\ee
In contrast to the requirements:
\be
       V_{eff}(\phi_0^2) = V'_{eff}(\phi_0^2) = 0,
                                          \lb{59}
\ee
producing the curve "1", let us consider now the joint solution of the
following equations:
\be
         V_{eff}(\phi_0^2) = V''_{eff}(\phi_0^2) = 0 .
                                         \lb{60}
\ee
The dashed curve "2" of Fig.3 represents the solution of
Eq.(\ref{60}). This curve is going very close to the maximum of the curve "1".
It is natural to assume that the position of the triple point A
coincides with this maximum and the corresponding deviation can be explained
by our approximate calculations. Taking into account such an assumption,
let us consider the border between the phase "conf.1" having the first
minimum at nonzero $\phi_0$  with $V_{eff}^{min} = c_1 < 0$
and the phase "conf.2 " which reveals two minima with the second minimum
being the deeper one and having $V_{eff}^{min}=c_2 < 0$.
This border (described by the curve "3" of Fig.3) was calculated in the
vicinity of the triple point A by means of Eqs.(\ref{57})
with $\phi_1$ and $\phi_2$ represented as $ \phi_{1,2} = \phi_0 \pm \epsilon$
with $\epsilon << \phi_0$. The result of such calculations gives the
following expression for the curve "3":
\be
  g^4_{run} = \frac 52 ( 5\lambda_{run} + 8 \pi^2) \lambda_{run} + 8\pi^4.
                                       \lb{61}
\ee
The piece of the curve "1" to the left of the point A describes the border
between the "Coulomb--like" phase and the phase "conf.1".
The right piece of the curve "1" along to the right of the point B
separates the "Coulomb" phase and the phase "conf.2". But
between the points A and B the phase transition border is going
slightly upper the curve "1". This deviation is very small and can't be
distinguished on Fig.3.

The numerical solution demonstrates that the triple point A
exists in the very neighborhood of the maximum of the curve (\ref{55})
and its position is approximately given by the following values:
\be
     \lambda_{(A)}\approx - \frac{4\pi^2}3\approx -13.4,
                                                     \lb{62}
\ee
\be
         g^2_{(A)} = g^2_{crit}|_{\mbox{for}\;
                      \lambda_{run}=\lambda_{(A)}}
                 \approx \frac{4\sqrt{2}}3{\pi^2}\approx 18.6.
                                                       \lb{63}
\ee
The triple point value of the electric and magnetic fine structure constant
follow from Eqs.(\ref{27}) and (\ref{63}):
\be
  \alpha_{(A)} = \frac{\pi}{g^2_{(A)}}\approx{0.17}, \qquad
\tilde \alpha_{(A)} = \frac {g^2_{(A)}}{4\pi}
                 \approx 1.48.       \lb{64}
\ee
The obtained result is very close to the Monte Carlo lattice result
(\ref{48}).

The phase diagram drawn in Fig.3 corresponds to the
validity of one--loop approximation (with accuracy of deviations
not more than $30\%$, see Ref.\ct{17}) in the region of parameters:
\be
0.17 \stackrel{<}{\sim }{\large \alpha, \tilde{\alpha}}
\stackrel{<}{\sim }1.5,         \lb{65}
\ee

It is necessary to note that the RGE for $\lambda_{run}$
indicates a slow convergence of the series over $\lambda$
(see Ref.\ct{33}) and the one--loop approximation is valid
for $\lambda_{run}$ up to $|\lambda|\stackrel{<}{\sim }30$ with accuracy
of deviations $<10\%$.

It is obvious that in our case both phases, "conf.1" and "conf.2",
have nonzero monopole condensate in the minima of the effective
potential, when $V_{eff}^{min}(\phi_{1,2}\neq 0) < 0$. By this reason, the
Abrikosov--Nielsen--Olesen (ANO) electric vortices
\ct{34},\ct{35} may exist in these both phases, which are
equivalent in the sense of the "string" formation.  If electric
charges are present in a model, they are placed at the ends of
the vortices--"strings" and therefore are confined.
The phase diagram of Fig.3 demonstrates the existence of the confinement
phase for $\alpha \ge \alpha_{(A)}\approx 0.17$.

The lattice investigations \ct{20},\ct{22} show that in the confinement phase
$\alpha (\beta)$ increases when $\beta = 1/e_0^2 \to 0$
(here $e_0$ is the the bare electric charge) and very slowly
approaches to its maximal value:
$\alpha_{max}=\frac {\pi}{12}\approx 0.26$ predicted in Ref.\ct{36}
(see also Ref.\ct{15}).

It is worthwhile mentioning that the confinement of monopoles
can be discribed by using duality.
The Higgs field $\Psi$, having the electric charge, is responsible
for this confinement.
The corresponding confinement phases for monopoles are absent
on the phase diagram of Fig.3. They can be described by the phase
diagram ($\lambda^e_{run}; e^2_{run}$).
The overall phase diagram is three-dimensional and is given by
$(\lambda^m_{run}; \lambda^e_{run}; g^2_{run})$ ($e^2_{run}$ and
$g^2_{run}$ are related by the Dirac relation).

The result (64) obtained in the framework of the Higgs
scalar monopole model gives the following prediction:
\be
           \alpha_{crit}^{-1} = {\alpha_{(A)}}^{-1}
                    \approx 6,                           \lb{66}
\ee
which is comparable with the MPM result (\ref{6}).

Although the one--loop approximation for the (improved) effective
potential does not give an exact coincidence
with the MPM prediction of the critical $\alpha$,
we see that, in general, the Higgs monopole model is very
encouraging for the AGUT--MPM. We have a hope that the two--loop
approximation corrections to the Coleman--Weinberg effective potential
will lead to the better accuracy in calculation of the phase transition
couplings.

The review of all existing results gives:

1)
\be
    \alpha_{crit}^{lat}\approx{0.20 \pm 0.015},\quad
    {\tilde \alpha}_{crit}^{lat}\approx{1.25 \pm 0.10}     \lb{67}
\ee
-- in the Compact QED with the Wilson lattice action \ct{20};

2)
\be
    \alpha_{crit}^{lat}\approx{0.204} \quad
    {\tilde \alpha}_{crit}^{lat}\approx{1.25}    \lb{68}
\ee
-- in the model with the Wilson loop action \ct{15};

3)
\be
   \alpha_{crit} \approx 0.1836,\quad \tilde \alpha_{crit} \approx 1.36
                                                  \lb{69}
\ee
-- in the Compact QED with the Villain lattice action \ct{22};

4)
\be
     \alpha_{crit} = \alpha_{(A)}\approx{0.17},\quad
     {\tilde \alpha}_{crit} = {\tilde \alpha}_{(A)}\approx 1.48
                                     \lb{70}
\ee
-- in the Higgs scalar monopole model (the present paper).

\vspace{0.1cm}

Hereby we see an additional arguments for our previously hoped (see
\ct{15} and \ct{16}) "approximate universality" of the first order
phase transition couplings: the fine structure constant (in the
continuum) is at the/a multiple point approximately the same one
independent of various parameters of the different (lattice, etc.)
regularization.

Recent investigations (L.V.Laperashvili, H.B.Nielsen, Bled Workshop,
Slovenia, 2000) show that monopoles are confined in the SM
up to the Planck scale. But they can exist in the AGUT.
Let us assume now the existence of monopoles at superhigh energies
$\mu \ge \mu_{mon}\stackrel{<}{\sim}\mu_{Pl}$. Then RGEs (\ref{45})
describing
also monopoles can lead to the Unification of all interactions
at the Planck scale giving the coincidence of
all $\alpha_{i,crit}$ at the point $\mu \sim \mu_{Pl}$.
Such a situation is shown in Fig.1 by dashed curves.
These investigations are in progress.

\newpage

{\bf Table I}: Best fit to conventional experimental data.
All masses are running masses at 1 GeV except the top qurk mass
$M_t$ which is the pole mass.

$~~~~~~~~~~~-----------------------  $

$~~~~~~~~~~~~~~~~~~~~~~~{\mbox {Fitted}}~~~~~~~~~~~~{\mbox{Experimental}} $

$~~~~~~~~~~~----------------------- $

$~~~~~~~~~~~~~m_u~~~~~~3.6~{\mbox{MeV}}~~~~~~~~~4~{\mbox{MeV}}$

$~~~~~~~~~~~~~m_d~~~~~~7.0~{\mbox{MeV}}~~~~~~~~~9~{\mbox{MeV}}$

$~~~~~~~~~~~~~m_e~~~~~~0.87~{\mbox{MeV}}~~~~~~~~0.5~{\mbox{MeV}}$

$~~~~~~~~~~~~~m_c~~~~~~1.02~{\mbox{GeV}}~~~~~~~~1.4~{\mbox{GeV}}$

$~~~~~~~~~~~~~m_s~~~~~~400~{\mbox{MeV}}~~~~~~~~~200~{\mbox{MeV}}$

$~~~~~~~~~~~~~m_{\mu}~~~~~~88~{\mbox{MeV}}~~~~~~~~~~105~{\mbox{MeV}}$

$~~~~~~~~~~~~~M_t~~~~~~192~{\mbox{GeV}}~~~~~~~~~180~{\mbox{GeV}}$

$~~~~~~~~~~~~~m_b~~~~~~~8.3~{\mbox{GeV}}~~~~~~~~~~6.3~{\mbox{GeV}}$

$~~~~~~~~~~~~~m_{\tau}~~~~~~1.27~{\mbox{GeV}}~~~~~~~~1.78~{\mbox{GeV}}$

$~~~~~~~~~~~~~V_{us}~~~~~~0.18~~~~~~~~~~~~~~~0.22$

$~~~~~~~~~~~~~V_{cb}~~~~~~0.018~~~~~~~~~~~~~~0.041$

$~~~~~~~~~~~~~V_{ub}~~~~~~0.0039~~~~~~~~~~~~0.0035$

$~~~~~~~~~~~-----------------------  $

\vspace{1cm}

{\bf Table II}: Best fit to conventional experimental data in the "new"
extended AGUT.
All masses are running masses at 1 GeV except the top qurk mass
$M_t$ which is the pole mass.

$~~~~~~~~~~~-----------------------  $

$~~~~~~~~~~~~~~~~~~~~~~~{\mbox {Fitted}}~~~~~~~~~~~~{\mbox{Experimental}} $

$~~~~~~~~~~~----------------------- $

$~~~~~~~~~~~~~m_u~~~~~~3.1~{\mbox{MeV}}~~~~~~~~~4~{\mbox{MeV}}$

$~~~~~~~~~~~~~m_d~~~~~~6.6~{\mbox{MeV}}~~~~~~~~~9~{\mbox{MeV}}$

$~~~~~~~~~~~~~m_e~~~~~~0.76~{\mbox{MeV}}~~~~~~~~0.5~{\mbox{MeV}}$

$~~~~~~~~~~~~~m_c~~~~~~1.29~{\mbox{GeV}}~~~~~~~~1.4~{\mbox{GeV}}$

$~~~~~~~~~~~~~m_s~~~~~~390~{\mbox{MeV}}~~~~~~~~~200~{\mbox{MeV}}$

$~~~~~~~~~~~~~m_{\mu}~~~~~~85~{\mbox{MeV}}~~~~~~~~~~105~{\mbox{MeV}}$

$~~~~~~~~~~~~~M_t~~~~~~179~{\mbox{GeV}}~~~~~~~~~180~{\mbox{GeV}}$

$~~~~~~~~~~~~~m_b~~~~~~~7.8~{\mbox{GeV}}~~~~~~~~~~6.3~{\mbox{GeV}}$

$~~~~~~~~~~~~~m_{\tau}~~~~~~1.29~{\mbox{GeV}}~~~~~~~~1.78~{\mbox{GeV}}$

$~~~~~~~~~~~~~V_{us}~~~~~~0.21~~~~~~~~~~~~~~~0.22$

$~~~~~~~~~~~~~V_{cb}~~~~~~0.023~~~~~~~~~~~~~~0.041$

$~~~~~~~~~~~~~V_{ub}~~~~~~0.0050~~~~~~~~~~~~0.0035$

$~~~~~~~~~~~-----------------------  $

\newpage

\begin{figure}[h]
\vspace{32mm}
\bc
\bp(0,0)
\put(0,0){\includegraphics{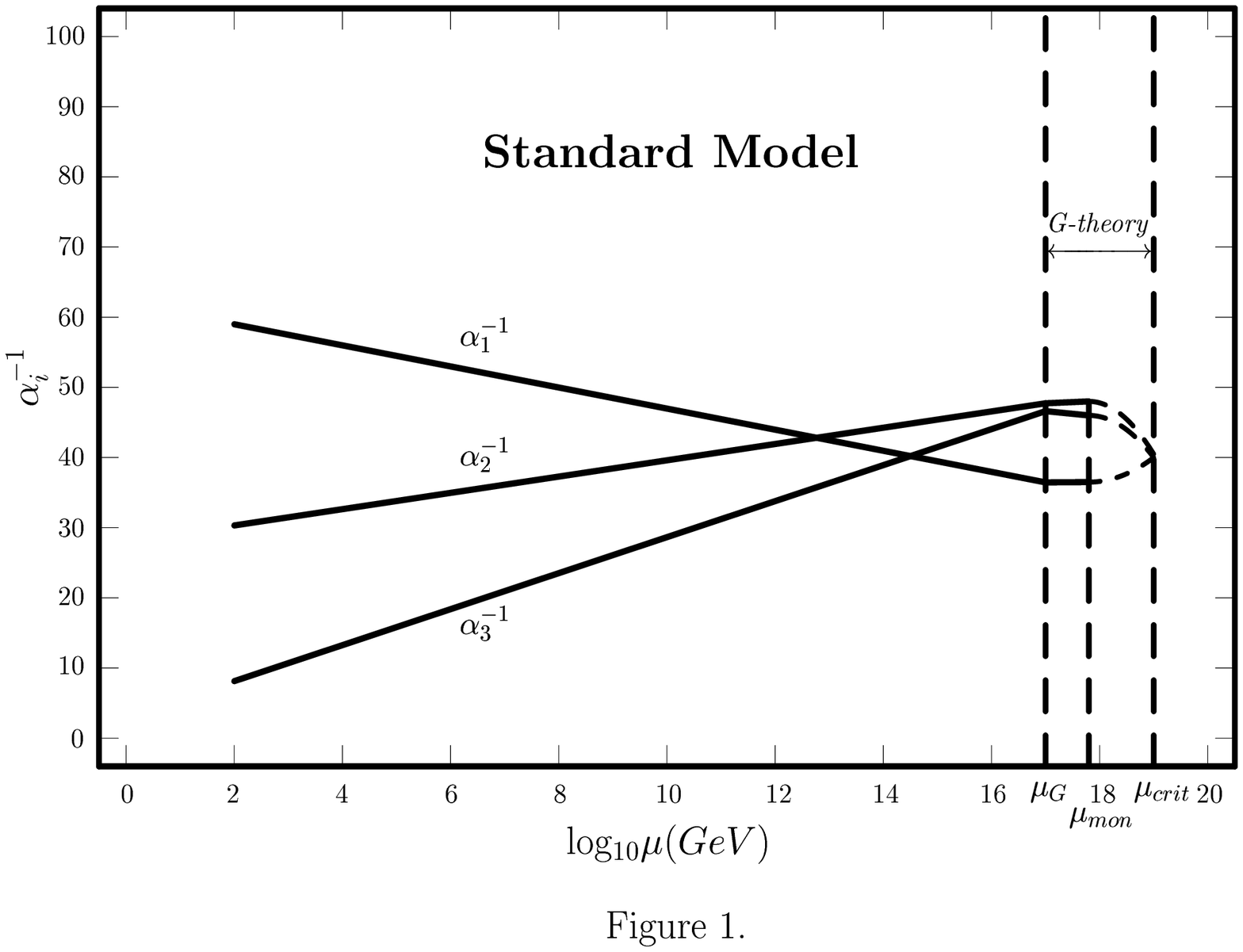}}
\ep
\ec
\end{figure}

\begin{figure}[h]
\vspace{37mm}
\bc
\bp(0,0)
\put(0,0){\includegraphics{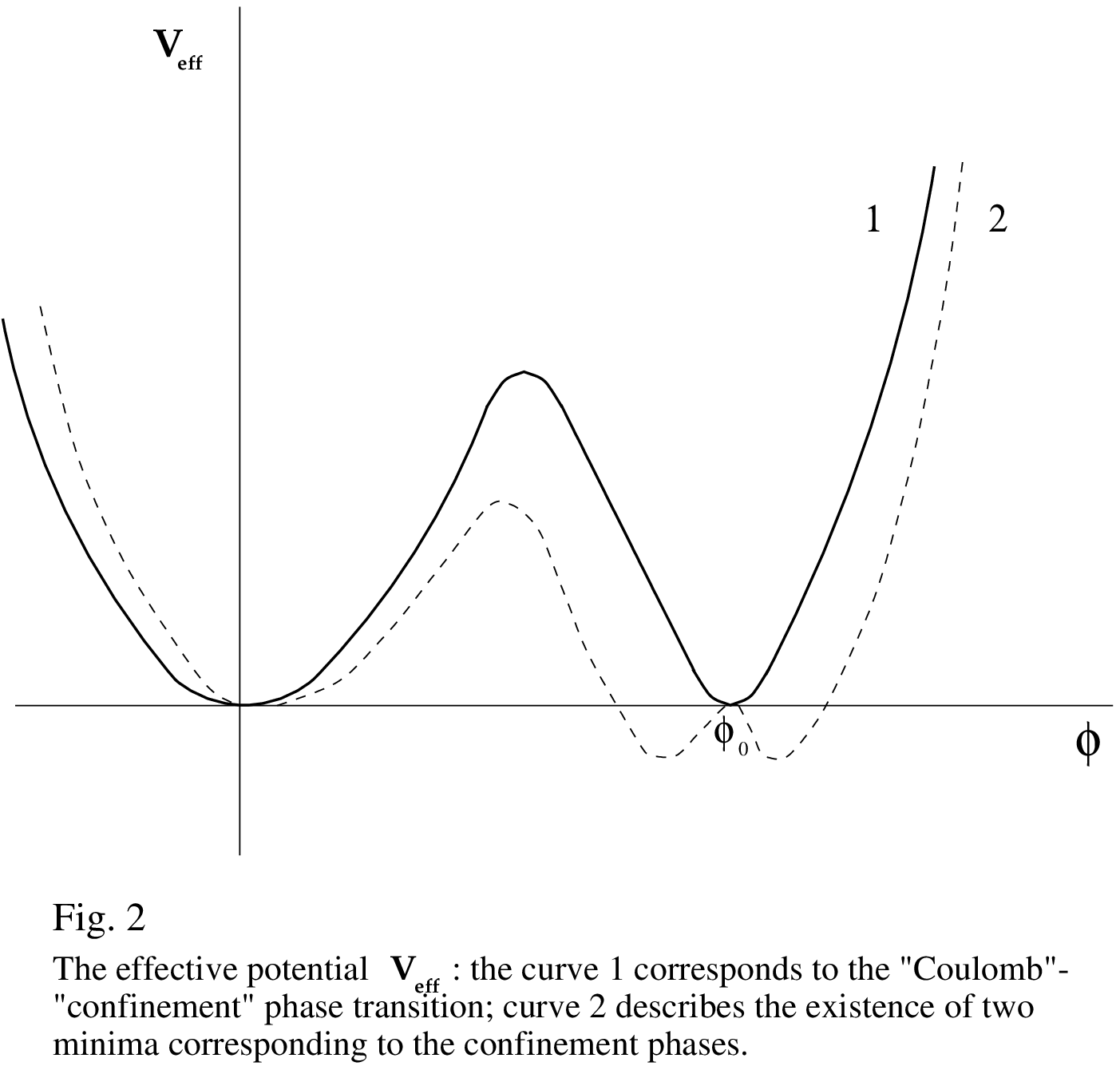}}
\ep
\ec
\end{figure}


\setcounter{figure}{2}

\begin{figure}[h]
\vspace{42mm}
\bc
\bp(0,0)
\put(0,0){\includegraphics{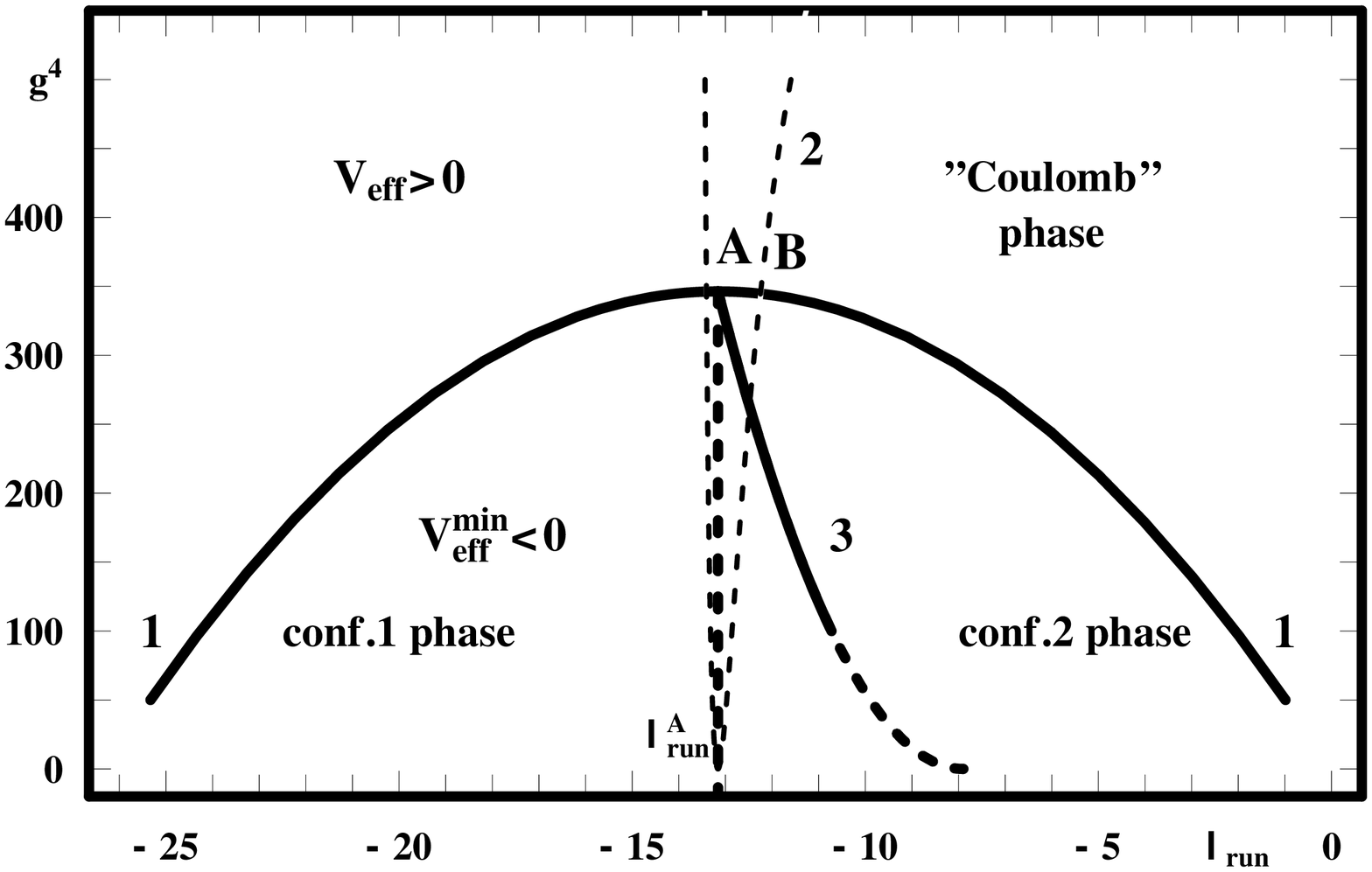}}
\ep
\ec
\caption{The phase diagram ($\lambda_{run};\; {\rm g}^4\equiv
g^4_{run}$)
corresponding to the Higgs monopole model shows the existence of a triple
point A $\bigl(\lambda_{(A)} \approx -13.4;\;{\rm g}^2_{(A)}\approx
18.6\bigr)$.}
\end{figure}

\end{document}